\documentclass[cits]{PoS}

\title{A two-year monitoring campaign of Supergiant Fast X-ray Transients with 
\emph{Swift} 
}

\ShortTitle{Monitoring SFXTs with {\it Swift}}

\author{ \speaker{P.\ Romano},$^a$ 
V.\ La Parola,$^a$ G.\ Cusumano,$^a$ S.\ Vercellone,$^a$ 
P.\ Esposito,$^b$ J.A.~Kennea,$^c$ D.N.\ Burrows,$^c$ H.A.\ Krimm,$^{de}$ C.\ Pagani,$^f$N.\ Gehrels,$^g$ \\
\llap{$^a$}INAF, Istituto di Astrofisica Spaziale e Fisica Cosmica, \\
         Via U.\ La Malfa 153, I-90146 Palermo, Italy \\
\llap{$^b$}INAF, Osservatorio Astronomico di Cagliari, \\
         localit\`a Poggio dei Pini, strada 54, I-09012 Capoterra, Italy\\
\llap{$^c$}Department of Astronomy and Astrophysics, Pennsylvania State  University, \\
         University Park, PA 16802, USA\\
\llap{$^d$}CRESST/Goddard Space Flight Center, Greenbelt, MD, USA\\
\llap{$^e$}Universities Space Research Association, Columbia, MD, USA\\
\llap{$^f$}Department of Physics \& Astronomy, University of Leicester, LE1 7RH, UK\\
\llap{$^g$}NASA/Goddard Space Flight Center, Greenbelt, MD 20771, USA\\
E-mail: \email{romano@ifc.inaf.it}   \hspace{1cm}         \href{http://www.ifc.inaf.it/sfxt/}{http://www.ifc.inaf.it/sfxt/ }

}

\abstract{
{\it Swift} is the only observatory which, due to its
unique fast-slewing capability and broad-band energy coverage,
can detect outbursts from Supergiant Fast X-ray Transients (SFXTs)
from the very beginning and study their evolution panchromatically. 
Thanks to its flexible observing scheduling, which makes monitoring 
cost-effective, {\it Swift} has also performed a campaign that covers 
all phases of the lives of SFXTs with a high sensitivity 
in the soft X--ray regime, where most SFXTs had not been
observed before. Our continued effort at monitorning SFXTs 
with 2--3 observations per week (1--2 ks) with the {\it Swift} X--Ray Telescope (XRT)
over their entire visibility period has just finished its second year. 
We report on our findings on the long-term properties of SFXTs, 
their duty cycle, and the new outbursts caught by {\it Swift} 
during the second year. 
}

\FullConference{8th INTEGRAL Workshop ``The Restless Gamma-ray Universe''\\
September 27-30 2010\\
Dublin Castle, Dublin, Ireland}

\begin{document}

\section{The \emph{Swift} SFXT Monitoring program}

Supergiant Fast X--ray Transients (SFXTs, \cite{Sguera2005}), the new class of 
High Mass X--ray Binaries (HMXBs) discovered by \emph{INTEGRAL}, 
are characterized by outbursts which are significantly shorter than 
those typical of Be/X-ray binaries, peak luminosities in the order of a few 
10$^{36}$~erg~s$^{-1}$, 
and a quiescent luminosity level of  $\sim 10^{32}$~erg~s$^{-1}$.
It is generally agreed that SFXTs are HMXBs with an
OB supergiant star companion to a neutron star (NS),
because their spectral properties are similar to those of accreting pulsars,  
even though a pulse period was measured in only a few of them.
The mechanisms responsible for the outbursts observed by \emph{INTEGRAL} 
involve either the structure of the wind from the supergiant companion 
\cite{zand2005,Walter2007,Negueruela2008,Sidoli2007}
or gated mechanisms (see \cite{Bozzo2008}). 

Thanks to its fast-slewing capability and its broad-band energy coverage,
{\it Swift} is the only observatory which  
can catch outbursts from these transients, observe them panchromatically 
from as short as 100\,s after their onset, and follow them 
as they evolve.
Furthermore, {\it Swift}'s flexible observing scheduling
makes a monitoring effort cost-effective. Thus, our campaign with
{\it Swift} has given SFXTs the first non-serendipitous attention in 
all phases of their lives with a high sensitivity
in the soft X-ray regime, where most SFXTs had not been
observed before. 

Our sample consists of 4 targets, IGR~J16479$-$4514, 
XTE~J1739--302/IGR~J17391$-$3021, 
IGR~J17544$-$2619, and AX~J1841.0$-$0536/IGR~J18410$-$0535, 
chosen among the 8 SFXTs known at the end of 2007, 
including the two prototypes of the class (XTE J1739--302, 
IGR  J17544$-$2619). 
During the second year of {\it Swift} observations, we monitored three targets, 
XTE J1739--302, IGR J17544$-$2619, and  IGR~J16479$-$4514. 
We obtained 2 or 3 observations per week per source, each 1\,ks long 
with the goal of systematically studying the outbursts, 
to monitor them during their evolution,
and for the very first time, 
to study the long term properties of SFXTs, in particular, 
the out-of-outburst states, and the quiescence.
This observing strategy was chosen to fit within the regular observing schedule
of $\gamma$-ray bursts (GRBs). 
Moreover, to ensure simultaneous narrow field instrument (NFI) 
data, the {\it Swift} Team enabled automatic rapid slews  
to these objects following detection of flares by the BAT,  
as is currently done for GRBs. 
We also requested target of opportunity (ToO)
observations whenever one of the sources showed interesting activity,
or following outbursts to better monitor the decay of the XRT light curve, 
thus obtaining a finer sampling of the light curves and allowing us to study all 
phases of the evolution of an outburst.

During the two years of monitorning we collected 
558 pointed XRT observations, for a total of 606\,ks of on-source exposure.
Table~\ref{igr10:tab:campaign} summarizes the campaign. 
The results on the long term X--ray properties outside the bright outbursts 
of our sample of SFXTs can be found in \cite{Sidoli2008:sfxts_paperI}, 
\cite{Romano2009:sfxts_paperV}, and \cite{Romano2010:sfxts_paperVI},
while the outbursts are analyzed in detail in 
\cite{Romano2008:sfxts_paperII}, 
\cite{Sidoli2009:sfxts_paperIII}, 
\cite{Sidoli2009:sfxts_paperIV} 
for IGR~J16479$-$4514 and the prototypical IGR~J17544$-$2619 
and XTE~J17391$-$302, respectively (also see Table~\ref{igr10:tab:campaign}).

\begin{table}
\begin{tabular}{lrrrrll}
\hline
\hline
Name &Campaign Dates &Obs.\ N.\    &{\it Swift}/XRT &Outburst & Outburst \\
     &               &             & Exposure (ks)  & Dates   & References\\
\hline
IGR~J16479--4514 &  2007-10-26--2009-11-01	& 144&	161&	2008-03-19&    \cite{Romano2008:sfxts_paperII} \\
                 &   	                        &    &     & 	2008-05-21&    \\ 
                 &   	                        &    &     & 	2009-01-29&     \cite{Romano2009:atel1920,LaParola2009:atel1929} \\ 
XTE~J1739--302 	 &  2007-10-27--2009-11-01	& 184&	206&	2008-04-08&	\cite{Romano2008:atel1466,Sidoli2009:sfxts_paperIII} \\
                 &   	                        &    &     & 	2008-08-13&	\cite{Romano2008:atel1659,Sidoli2009:sfxts_paperIV} \\
                 &   	                        &    &     & 	2009-03-10&	\cite{Romano2009:atel1961,Romano2010:sfxts_paperVI} \\
IGR~J17544--2619 &  2007-10-28--2009-11-03	& 142&	143&	2007-11-08&	\cite{Krimm2007:ATel1265} \\
  		 &    	                        &    &     & 	2008-03-31&     \cite{Sidoli2008:atel1454,Sidoli2009:sfxts_paperIII}\\
  		 &   	                        &    &     & 	2008-09-04& 	\cite{Romano2008:atel1697,Sidoli2009:sfxts_paperIV}\\
 		 &   	                        &    &     & 	2009-03-15& 	\cite{Krimm2009:atel1971,Romano2010:sfxts_paperVI} \\
 		 &   	                        &    &     & 	2009-06-06& 	\cite{Romano2009:atel2069,Romano2010:sfxts_paperVI} \\
AX~J1841.0--0536 &  2007-10-26--2008-11-15	&  88&	 96&	none      &      \\
\hline
\end{tabular}
\caption{The {\it Swift} monitoring campaign. The outburst dates refer to the outbursts that occurred during the monitoring. }
\label{igr10:tab:campaign}
\end{table}

\begin{figure}[t]
		\hspace{+1.5truecm}
 \includegraphics[width=0.6\textwidth,height=0.5\textheight,angle=0]{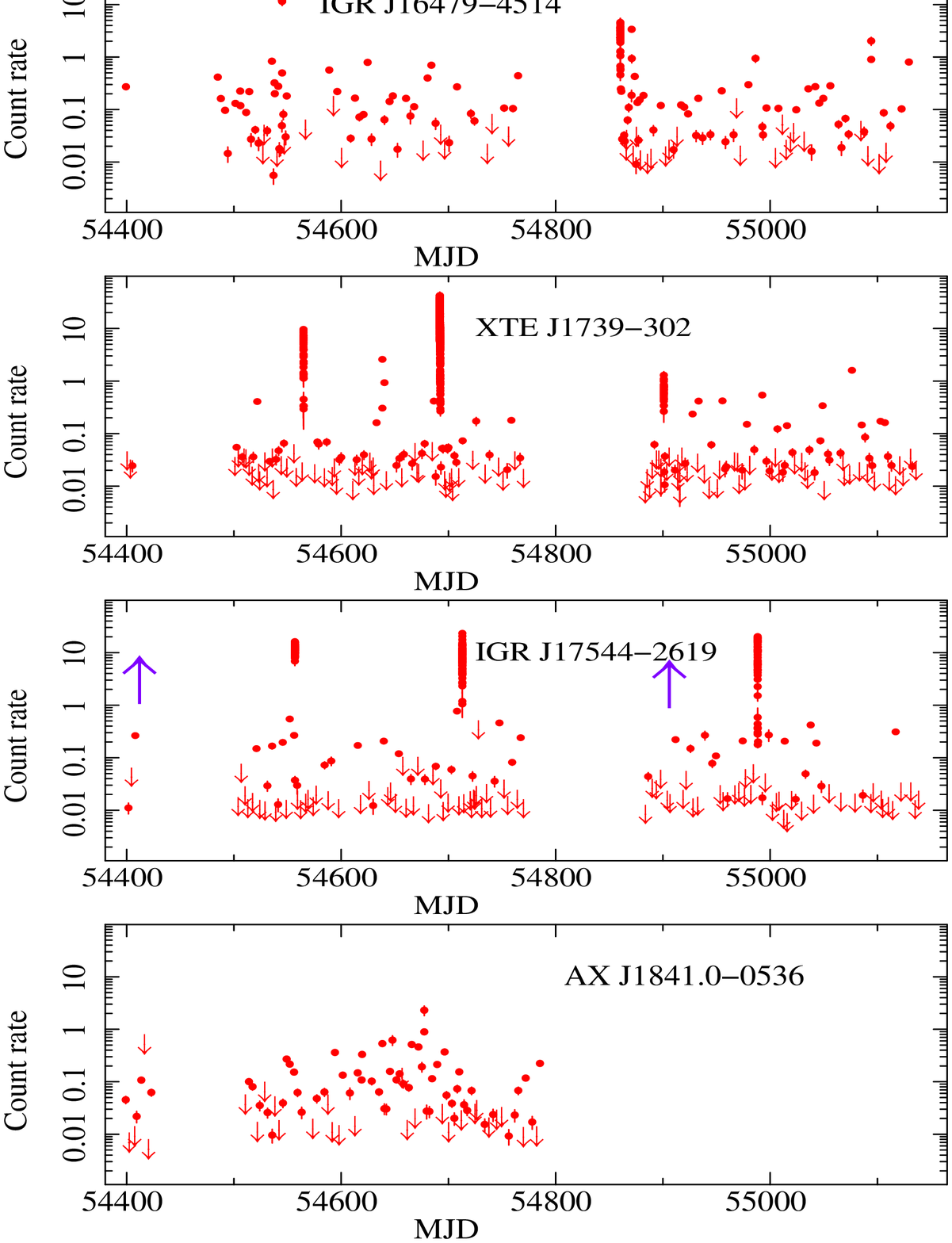}
		\vspace{-1.2truecm}
\caption{ {\it Swift}/XRT light curves of our sample in the
  0.2--10\,keV energy range, between 2007 October 26 and 2009
  November 3. The light curves are background subtracted, 
  corrected for  pile-up (when required), PSF losses, and vignetting. 
Each point refers to the average flux observed
during each observation performed with XRT, except for outbursts 
(Table~1) where the data were binned to 
include at least 20 source counts per time bin to best represent the 
dynamical range. Downward-pointing arrows are 3-$\sigma$ upper limits, 
upward pointing arrows mark either outbursts that  
XRT could not observe because the source was Sun-constrained,
or BAT Transient Monitor bright flares. 
AX~J1841.0$-$0536 was only observed during the first year. 
}
\label{igr10:fig:xrtlcvs}
\end{figure}

           \section{Light curves and inactivity duty cycle}
The 0.2--10\,keV XRT light curves collected from 2007 October 26 to  2009 November 3,
are shown in Fig.~\ref{igr10:fig:xrtlcvs}. 
Since our monitoring is a casual sampling of the light curves at a 
resolution of $\sim 3$--4\,d  over a $> 2$ yr baseline  (1\,yr for AX~J1841.0$-$0536), 
we infer that these sources spend 3--5\% of the total time in bright outbursts.

We address the issue of the percentage of time each source spends in each flux state.
We considered the following three states, 
{\it i)} BAT-detected outbursts, 
{\it ii)} intermediate states (firm detections excluding outbursts),  
{\it iii)} `non detections' (detections with a significance below 3$\sigma$). 
From the latter state we excluded all observations that had a net exposure below 900\,s
[corresponding to 2--10\,keV flux limits that vary between 1 and 3$\times 10^{-12}$ erg cm$^{-2}$ s$^{-1}$ 
(3$\sigma$), depending on the source, see \cite{Romano2009:sfxts_paperV}]. This was done because 
{\it Swift} is a GRB-chasing mission and several observations were interrupted by GRB events; therefore 
the consequent non detection may be due to the short exposure, and not exclusively to the source 
being faint. 

The duty cycle of {\it inactivity} is defined  \cite{Romano2009:sfxts_paperV}  as 
the time each source spends {\it undetected} down to a flux limit of 
1--3$\times10^{-12}$ erg cm$^{-2}$ s$^{-1}$,  
$
{\rm IDC}= \Delta T_{\Sigma} / [\Delta T_{\rm tot} \, (1-P_{\rm short}) ] \, ,     
$
where  
$\Delta T_{\Sigma}$ is sum of the exposures accumulated in all observations, 
   each in excess of 900\,s, where only a 3-$\sigma$ upper limit was achieved,  
$\Delta T_{\rm tot}$ is the total exposure accumulated (Table~\ref{igr10:tab:campaign}), and 
$P_{\rm short}$ is the percentage of time lost to short observations  
(exposure $<900$\,s, Table~\ref{igr10:tab:dutycycle}, column 3). 
We obtain that ${\rm IDC} = 19, 28, 39, 55$\,\%, 
for IGR~J16479$-$4514, AX~J1841.0$-$0536, XTE~J1739--302, and IGR~J17544$-$2619, respectively
(Table~\ref{igr10:tab:dutycycle}, column 4), with an estimated error 
of $\sim 5\,\%$.

\begin{table}
\begin{tabular}{lcccc}
\hline
\hline
Name &$\Delta T_{\Sigma}$  & $P_{\rm short}$ &  IDC  & Rate$_{\Delta T_{\Sigma}}$   \\           
                &(ks) & (\%) &  (\%) &   ($10^{-3}$counts s$^{-1}$)  \\  
\hline
IGR~J16479$-$4514   & 29.7 &3  & 19 & $3.1\pm0.5$ \\
XTE~J1739$-$302     & 71.5 &10 & 39 & $4.0\pm0.3$ \\
IGR~J17544$-$2619   & 69.3 &10 & 55 & $2.2\pm0.2$ \\
AX~J1841.0$-$0536   & 26.6 &3  & 28 & $2.4\pm0.4$ \\
\hline
\end{tabular}
\caption{Duty cycle of inactivity. }
\label{igr10:tab:dutycycle}
\end{table}

           \section{Spectroscopy}

\begin{figure}[t]
 		\hspace{-1.truecm}
                \centerline{\includegraphics[width=17cm,angle=0]{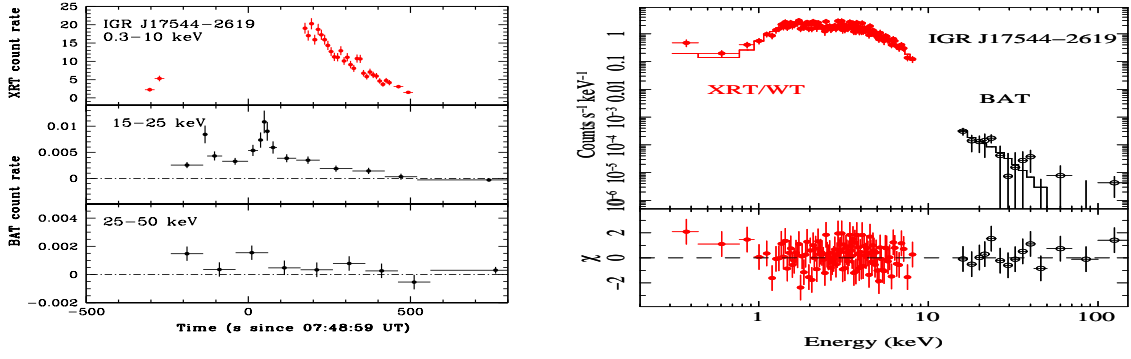}	}
 		\vspace{-1.truecm}
                 \caption{{\bf Left}: XRT (red) and BAT (black) light curves of the 2009 
                   June 6 outburst of IGR~J17544$-$2619  
		in units of count s$^{-1}$ and count s$^{-1}$ detector$^{-1}$, respectively. 
                The XRT data preceding the outburst were collected 
                as a pointed observation, part of our monitoring program. 
                {\bf Right}: Data from the XRT/WT spectrum and simultaneous BAT spectrum 
                fit with a {\sc cutoffpl} model, and residuals in units of standard deviations.
		}
                 \label{igr10:fig:outburst}
       \end{figure}

Simultaneous observations with XRT and BAT allowed us to perform 
broad-band spectroscopy of outbursts of SFXTs
from 0.3 kev to 100--150\,keV. 
This yielded particularly valuable information, 
because of the shape of the spectrum, a hard power law below 10\,keV
with a high-energy cutoff at 15--30\,keV. 
Therefore {\it Swift} is the ideal observatory
to study their spectrum: BAT constrains the hard-X spectral properties 
(to compare with the most popular accreting NS models) while XRT gives us a measurement of
the absorption, which is quite high in these objects \cite{Walter2006}, 
often well above the Galactic value. 
As an example, we report the properties of the 2009 June 6 outburst of IGR~J17544$-$2619
(Fig.~\ref{igr10:fig:outburst}) for which simultaneous BAT and XRT data were collected. 
An absorbed power-law model is inadequate so we considered 
an absorbed power-law model with a high energy cut-off 
($N_{\rm H}=1.0_{-0.3}^{+0.2}\times10^{22}$ cm$^{-2}$,  
$\Gamma=0.6_{-0.4}^{+0.2}$, E$_{\rm c}=3_{-1}^{+1}$ keV,
E$_{\rm f}=8_{-3}^{+4}$ keV,  $\chi^{2}_{\nu}$/dof$=0.92/115$)      
and an absorbed  power-law model with an exponential cutoff
($N_{\rm H}=1.0_{-0.2}^{+0.3}\times10^{22}$ cm$^{-2}$,  
$\Gamma=0.4_{-0.3}^{+0.3}$, E$_{\rm c}=7_{-2}^{+4}$ keV, $\chi^{2}_{\nu}$/dof$=0.94/116$),   
models typically used to describe the X--ray emission from 
accreting NS in HMXBs.  

On the other hand, our {\it Swift}/XRT monitoring campaign has demonstrated for the
first time that X--ray emission from SFXTs is still present {\it outside}
the bright outbursts, although at a much lower level 
\cite{Sidoli2008:sfxts_paperI,Romano2009:sfxts_paperV,Romano2010:sfxts_paperVI}.
Spectral fits performed in the 0.3--10\,keV energy band by adopting simple models
such as an absorbed power law or a blackbody (more complex
models were not required by the data) result in hard power law photon
indices (always in the range ~0.8--2.1) or in hot blackbodies
($kT_{\rm BB}\sim 1$--2\,keV).

     \begin{figure}[t]
 		\vspace{-4.5truecm}
 		\hspace{-0.5truecm}
                \centerline{\includegraphics[width=16cm,height=12cm,angle=0]{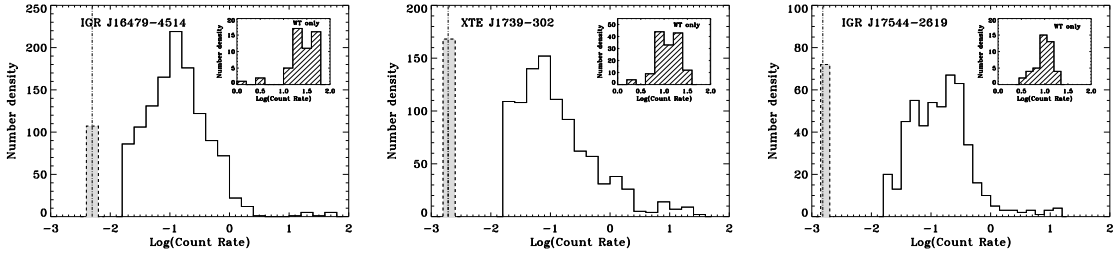}	}
		\vspace{-4truecm}
                 \caption{Distribution of the count rates when the XRT light curves are binned at 100\,s,
for the three sources monitored for two years.  
The vertical lines correspond to the background. The hashed histograms are points which are consistent with 
a zero count rate. The insets show the subset of WT data only, binned at 20\,s.
		}
                \label{igr10:fig:histos}
       \end{figure}

           \section{Long term properties}

Figure~\ref{igr10:fig:histos} shows the distributions of the observed count rates 
after removal of the observations where a detection was not achieved. 
A roughly Gaussian shape is observed, 
with a broad peak at $\approx 0.1$ counts s$^{-1}$, 
and a clear cut at the detection limit for 100\,s at the low end.
In particular, when the distributions are fit with a Gaussian function 
we find that their means are 
0.12  counts s$^{-1}$  (IGR~J16479$-$4514), 
0.06  counts s$^{-1}$  (XTE~J1739--302), and 
0.13  counts s$^{-1}$ (IGR~J17544$-$2619). Therefore, 
the most probable flux level at which a random observation
will find these sources, when detected, is 
$3 \times 10^{-11}$, $9\times 10^{-12}$, and $1 \times 10^{-11}$ erg cm$^{-2}$ s$^{-1}$ (unabsorbed 2--10\,keV,
i.e. luminosities of $\sim 8\times 10^{34}$, $8\times 10^{33}$, and
 $2\times 10^{34}$ erg s$^{-1}$), respectively. 

X-ray variability is observed at all timescales and intensities
we can probe. Superimposed on the day-to-day variability is intra-day flaring 
which involves variations up to one order of magnitude that can occur down to 
timescales as short as 1\,ks. These cannot be accounted for 
by accretion from a homogeneous wind, but can be naturally explained 
by the accretion of single clumps composing the donor wind. 
If, for example, we assume that each of these short flares is caused 
by the accretion of a single clump onto the NS \cite{zand2005},
then its mass can be estimated \cite{Walter2007} as 
$M_{\rm cl}= 7.5\times 10^{21} \,\, (L_{\rm X, 36}) (t_{\rm fl, 3{\rm ks}})^{3}$ g,
where $L_{\rm X, 36}$ is the average X-ray luminosity in units of $10^{36}$ erg s$^{-1}$,
$t_{\rm fl, 3{\rm ks}}$ is the duration of the flares in units of 3\,ks. 
We can confidently identify flares down to a count rate in the order of 
0.1\,counts s$^{-1}$; 
these correspond to luminosities in the order of 2--$6\times10^{34}$ erg s$^{-1}$, which yield
 $M_{\rm cl} \sim 0.3$--$2\times10^{19}$ g. 
These masses are about those expected \cite{Walter2007} to be
responsible for short flares, below the \emph{INTEGRAL}  detection threshold and which,
if frequent enough, may significantly contribute to the mass-loss rate.

{\bf Grants:} ASI I/088/06/0, NASA NAS5-00136.

\end{document}